\documentstyle[12pt,epsfig]{article}
\advance\textheight by \topskip
\begin{document}
\tolerance=100000
\thispagestyle{empty}
\setcounter{page}{0}

\newcommand{\be}{\begin{equation}}
\newcommand{\ee}{\end{equation}}
\newcommand{\br}{\begin{eqnarray}}
\newcommand{\er}{\end{eqnarray}}
\newcommand{\ba}{\begin{array}}
\newcommand{\ea}{\end{array}}
\newcommand{\bi}{\begin{itemize}}
\newcommand{\ei}{\end{itemize}}
\newcommand{\bn}{\begin{enumerate}}
\newcommand{\en}{\end{enumerate}}
\newcommand{\bc}{\begin{center}}
\newcommand{\ec}{\end{center}}
\newcommand{\ul}{\underline}
\newcommand{\ol}{\overline}
\newcommand{\ar}{\rightarrow}
\newcommand{\sm}{${\cal {SM}}$}
\newcommand{\susy}{{{SUSY}}}
\newcommand{\Dir}{\kern -6.4pt\Big{/}}
\newcommand{\Dirin}{\kern -12.4pt\Big{/}\kern 4.4pt}
\newcommand{\DGir}{\kern -6.0pt\Big{/}}
\def\pp{\ifmmode{pp} \else{$pp$} \fi}
\def\CC{\ifmmode{{\it C.C.}} \else{$\mbox{\it C.C.}$} \fi}
\newcommand{\bqbqH}{\ifmmode{bU\ar bDH^+} 
                   \else{$bU\ar bDH^+$}\fi}
\newcommand{\bqbqW}{\ifmmode{bU\ar bDW^+} 
                   \else{$bU\ar bDW^+$}\fi}
\newcommand{\qqbbW}{\ifmmode{U\bar D\ar b\bar bW^+} 
                   \else{$U\bar D\ar b\bar bW^+$}\fi}
\newcommand{\bqbqtn}{\ifmmode{bU\ar bD\tau^+\nu_\tau} 
                   \else{$bU\ar bD\tau^+\nu_\tau$}\fi}
\newcommand{\qqbbtn}{\ifmmode{U\bar D\ar b\bar b\tau^+\nu_\tau} 
                   \else{$U\bar D\ar b\bar b\tau^+\nu_\tau$}\fi}
\newcommand{\Htn}{\ifmmode{H^\pm\ar \tau\nu_\tau}
                   \else{$H^\pm\ar \tau\nu_\tau$}\fi}
\newcommand{\Wtn}{\ifmmode{W^\pm\ar \tau\nu_\tau}
                   \else{$W^\pm\ar \tau\nu_\tau$}\fi}
\def\mssm{\ifmmode{{\cal {MSSM}}}\else{${\cal {MSSM}}$}\fi}
\def\bbtowh{\ifmmode{b\bar b\ar W^\pm H^\mp}\else
           {${b\bar b\ar W^\pm H^\mp}$}\fi}
\def\ggtowh{\ifmmode{gg\ar W^\pm H^\mp}\else{${gg\ar W^\pm H^\mp}$}\fi}
\def\wpmhmp{\ifmmode{W^\pm H^\mp}\else{${W^\pm H^\mp}$}\fi}
\def\MH{\ifmmode{{M_{H}}}\else{${M_{H}}$}\fi}
\def\Mh{\ifmmode{{M_{h}}}\else{${M_{h}}$}\fi}
\def\MA{\ifmmode{{M_{A}}}\else{${M_{A}}$}\fi}
\def\MHpm{\ifmmode{{M_{H^\pm}}}\else{${M_{H^\pm}}$}\fi}
\def\Hpm{\ifmmode{{{H^\pm}}}\else{${{H^\pm}}$}\fi}
\def\tb{\ifmmode{\tan\beta}\else{$\tan\beta$}\fi}
\def\ctb{\ifmmode{\cot\beta}\else{$\cot\beta$}\fi}
\def\ta{\ifmmode{\tan\alpha}\else{$\tan\alpha$}\fi}
\def\cta{\ifmmode{\cot\alpha}\else{$\cot\alpha$}\fi}
\def\tba{\ifmmode{\tan\beta=1.5}\else{$\tan\beta=1.5$}\fi}
\def\tbb{\ifmmode{\tan\beta=30}\else{$\tan\beta=30.$}\fi}
\def\cab{\ifmmode{c_{\alpha\beta}}\else{$c_{\alpha\beta}$}\fi}
\def\sab{\ifmmode{s_{\alpha\beta}}\else{$s_{\alpha\beta}$}\fi}
\def\cba{\ifmmode{c_{\beta\alpha}}\else{$c_{\beta\alpha}$}\fi}
\def\sba{\ifmmode{s_{\beta\alpha}}\else{$s_{\beta\alpha}$}\fi}
\def\ca{\ifmmode{c_{\alpha}}\else{$c_{\alpha}$}\fi}
\def\sa{\ifmmode{s_{\alpha}}\else{$s_{\alpha}$}\fi}
\def\cb{\ifmmode{c_{\beta}}\else{$c_{\beta}$}\fi}
\def\sb{\ifmmode{s_{\beta}}\else{$s_{\beta}$}\fi}

\def\Ord{\buildrel{\scriptscriptstyle <}\over{\scriptscriptstyle\sim}}
\def\OOrd{\buildrel{\scriptscriptstyle >}\over{\scriptscriptstyle\sim}}
\def\pl #1 #2 #3 {{\it Phys.~Lett.} {\bf#1} (#2) #3}
\def\np #1 #2 #3 {{\it Nucl.~Phys.} {\bf#1} (#2) #3}
\def\zp #1 #2 #3 {{\it Z.~Phys.} {\bf#1} (#2) #3}
\def\pr #1 #2 #3 {{\it Phys.~Rev.} {\bf#1} (#2) #3}
\def\prep #1 #2 #3 {{\it Phys.~Rep.} {\bf#1} (#2) #3}
\def\prl #1 #2 #3 {{\it Phys.~Rev.~Lett.} {\bf#1} (#2) #3}
\def\mpl #1 #2 #3 {{\it Mod.~Phys.~Lett.} {\bf#1} (#2) #3}
\def\rmp #1 #2 #3 {{\it Rev. Mod. Phys.} {\bf#1} (#2) #3}
\def\sjnp #1 #2 #3 {{\it Sov. J. Nucl. Phys.} {\bf#1} (#2) #3}
\def\cpc #1 #2 #3 {{\it Comp. Phys. Comm.} {\bf#1} (#2) #3}
\def\xx #1 #2 #3 {{\bf#1}, (#2) #3}
\def\preprint{{\it preprint}}

\begin{flushright}
{Cavendish-HEP-98/13}\\ 
{RAL-TR-1998-063}\\ 
{September 1998\hspace*{.5 truecm}}\\
\end{flushright}

\vspace*{\fill}

\begin{center}
{\Large \bf 
The phenomenology of \wpmhmp\ production\\[0.25cm]
at the Large Hadron Collider}\\[1.5cm]
{\large Stefano Moretti$^{a}$ and Kosuke Odagiri$^{b}$}\\[0.4 cm]
{\it a) Rutherford Appleton Laboratory,}\\
{\it Chilton, Didcot, Oxon OX11 0QX, UK.}\\[0.25cm]
{\it b) Cavendish Laboratory, University of Cambridge,}\\
{\it Madingley Road, Cambridge CB3 0HE, UK.}\\[0.5cm]
\end{center}
\vspace*{\fill}

\begin{abstract}
{\noindent 
\small
Barrientos Bendez\'u and Kniehl \cite{BBK} recently suggested that
\wpmhmp\ associated production may be a useful channel in the search for
the elusive heavy charged Higgs bosons of the 2 Higgs Doublet Model at the
Large Hadron Collider. We investigate the phenomenology of this mechanism 
in the Minimal Supersymmetric Standard Model, with
special attention paid to the most likely heavy Higgs decay,
$H^\mp\rightarrow tb\rightarrow b\bar b W^\mp$, 
and to the irreducible background from top pair production. We find that 
the semi-leptonic signature `$b\bar b W^+W^-$ $\rightarrow$ 
$b\bar b~jj~\ell~+~\mbox{missing~momentum}$' is dominated by top-antitop
events, which overwhelm the charged Higgs signal over the
heavy mass range that can be probed at the CERN collider.}
\end{abstract}
\vskip5.0cm
\hrule
\vskip0.25cm
\noindent
Electronic mails: moretti@v2.rl.ac.uk,~odagiri@hep.phy.cam.ac.uk.

\vspace*{\fill}
\newpage

\section*{1. Introduction}

In Ref.~\cite{BBK}, Barrientos Bendez\'u and Kniehl pointed out 
that the processes
\be\label{bb}
\bbtowh\ 
\ee
and 
\be\label{gg}
\ggtowh\ 
\ee
can be used in the search for the heavy
($M_{H^\pm}>m_t+m_b$) charged Higgs bosons $H^\pm$ of the 2 Higgs
Doublet Model (2HDM) at the Large Hadron Collider (LHC).

This result is particularly welcome, since it has been remarked in several
occasions (see Ref.~\cite{ep} for an overview) that at present it is not
at all certain that such particles can be detected at the LHC, even in the
Minimal Supersymmetric Standard Model (MSSM), if the typical energy scale of
Supersymmetry (SUSY) is much greater than the charged Higgs mass. 

Previous studies have shown that, if
$M_{H^\pm}> m_t+m_b$, the chances of $H^\pm$ detection at the LHC are
reliant only on two production mechanisms: the subprocesses $bg\ar tH^\pm$
\cite{guide81} and $bq\ar bq'H^\pm$ \cite{bW} and provided that
$\MHpm\Ord300-400$ GeV \cite{guide}. These channels generally have poor
signal-to-background ratios, as the event signatures always involve a
large number of jets, which is the typical noise of a hadron-hadron
machine.

When $H^\pm$'s are heavy, for $M_{\mathrm{SUSY}}\gg\MHpm$, 
they decay almost exclusively to $\bar{b}t
(b\bar{t})$ \cite{BRs}. In addition, hadronic signatures of $W^\pm$ bosons
produced in top decays are normally selected in order to allow for the $\Hpm$
mass reconstruction. Therefore, at first sight, it appears that
the signals (\ref{bb})--(\ref{gg}) advocated in Ref.~\cite{BBK} as a
useful source of Higgs events may be swamped by the irreducible background
from top pair production,
\be\label{tt}
q\bar q\ar t\bar t\qquad\qquad{\mbox{and}}\qquad\qquad gg\ar t\bar t,
\ee
with subsequent decay through the intermediate
state\footnote{Alternatively, in a narrow window in $\MHpm$ and only at
low $\tan\beta$, the charged Higgs bosons can decay to $W^\pm h$ pairs, where
$h$ represents the lightest Higgs boson. Although we will not treat this
case here, we note that even in this channel the final state is identical
to that of top-antitop, as $h\ar b\bar b$ is dominant over most of the SUSY
parameter space.} $b\bar bW^+W^-$. In order to understand whether this is
the case, we made a detailed signal-to-background analysis 
and found that top-antitop production and decay indeed overwhelms 
the new Higgs signal. 

The plan of this paper is as follows. In the next Section we describe the
details of the calculation and list the parameter values adopted. Section
3 is devoted to the discussion of results. We present our conclusions in
Section 4.

\section*{2. Calculation}

We generated the signal cross sections by using the formulae of
Ref.~\cite{BBK}. However, for phenomenological analyses we need 
to supplement those expressions in several ways. 

Firstly, their matrix elements (MEs) do
not include the $W^\pm$ boson decay and thus carry no information on the
angular distributions of the fermion pair it produces. For the $H^\pm$
this does not matter, since scalar particles decay isotropically. However,
even in this case one has to provide the correct kinematics for the decay
sequence $\Hpm\ar tb\ar b\bar b W^\pm\ar b\bar b jj$ (where $j$ represents a 
light quark jet produced in the $W^\pm$ decay), which we have done by 
computing the exact ME constructed by means of the {\tt HELAS} subroutines
\cite{HELAS}. As for the $W^\pm$ decay in the production channels 
(\ref{bb}) and (\ref{gg}), we have re-evaluated their MEs with the 
additional insertion of the $W^\pm$ boson decay
currents. For the $b\bar{b}$ fusion case, the actual expression is the same 
as for 
the $bq\ar bq'H^\pm$ process calculated in \cite{bW} and recently modified in
\cite{herwigMSSM}, but with some leg crossings. For the $gg$ fusion case,
the result is simply the replacement
\begin{equation}\label{ggnew}
\lambda(s,M_{W^\pm}^2,M_{H^\pm}^2)\ar4g^2M_{W^\pm}^2|{\mathcal G}_{W^\pm}|^2
[2(p_1\cdot p_{H^\pm})(p_2\cdot p_{H^\pm})-(p_1\cdot p_2)M_{H^\pm}^2]
\end{equation}
in equation (8) of \cite{BBK}, where $g^2=4\pi\alpha_{em}/\sin\theta_W^2$,
$|{\mathcal G}_{W^\pm}|^2=
 [(p_{W^\pm}^2-M_{W^\pm}^2)^2+(\Gamma_{W^\pm}M_{W^\pm})^2]^{-1}$, with
$p_{H^\pm}$, $p_{W^\pm}$, $p_{1}$ and $p_{2}$ the four-momentum of
the $H^\pm$, $W^\pm$, first and second lepton, say $\ell$ and $\nu_\ell$,
from the gauge boson decay, respectively.

Secondly, their MEs for the $gg$ fusion processes do not involve
squark loops, this preventing one from studying possible effects of
the SUSY partners of ordinary quarks, when their mass is below
the TeV scale. For example, these corrections are expected to be sizable in 
the MSSM, which is adopted here as the default SUSY framework. In this respect, 
we have modified the $gg$ triangle formula of Ref.~\cite{BBK} for the case of 
intermediate neutral Higgs production (top graph in Fig. 2 there),
by inserting the well known \cite{guide,xggh,SDGZ} squark loop
terms \cite{guide}.

Conversely, we have not included here the contribution of the 
box diagrams (bottom two graphs of Fig. 2 in Ref.~\cite{BBK}), 
for which the authors of that paper found no compact expression. 
According to their curves in Fig.~5 \cite{BBK}, this should result in an 
overestimate of the total cross section of subprocess (\ref{gg}), as the
triangle and box diagrams interfere destructively. The overall effect is
however negligible at large $\tan\beta$, 
the regime where the $W^\pm H^\mp$ cross section is larger
(whereas, for $\tba$, it can at times be more than a factor of two:
see Fig.~3 of Ref.~\cite{BBK}). In addition,
this is particularly true for $\MHpm\gg m_t$, the mass interval with which 
this paper is concerned. Indeed, in most of our plots we will concentrate
on that portion of the MSSM parameter space, for which we expect our results
to be reliable.

Finally, all our calculations for the signal were tested against the original 
cross sections of \cite{BBK}, and also using MadGraph  \cite{tim} for the case 
\bbtowh.

For the background we have assumed that the QCD contribution is reducible
by cuts on the reconstructed top and $W^\pm$ masses. Therefore, we studied
the top pair background only, which is anyhow the dominant component of
the final state $b\bar b U\bar D \ell\bar\nu_\ell$ where $U$ and $D$ refer
to up- and down-type massless quarks and $\ell=e$ or $\mu$. We have
considered both $q\bar q \ar t\bar t\ar b\bar b U\bar D \ell\bar\nu_\ell$
and $gg\ar t\bar t\ar b\bar b U\bar D \ell\bar\nu_\ell$, which we
generated at leading order using the HELAS library \cite{HELAS}. The
outputs of the corresponding code agree with the results given in
Ref.~\cite{gattoevolpe} when the $W^\pm$'s are on the mass shell.
In fact, notice that finite widths effects of top quarks,
gauge and charged Higgs bosons have been taken into account here.

Concerning the values of the various MSSM parameters entering the
computation of the signal processes (\ref{bb})--(\ref{gg}) (and,
marginally, the MSSM top width), we proceeded as follows. First,
we produced the masses and the widths of the Higgs bosons by means of
the two loop relations of Ref.~\cite{two-loop} (see also
\cite{new-two-loop}). To simplify the discussion, we have assumed a
universal soft Supersymmetry--breaking mass $m_{\tilde u}^2=m_{\tilde
d}^2\equiv m_{\tilde q}^2$, and negligible mixing in the stop and sbottom
mass matrices, $A_t=A_b=\mu=0$. Second, the MSSM Higgs widths were
generated using the program {\tt HDECAY} \cite{HDECAY}, which in turn uses
mass relations at the same perturbative level. Squark masses entering the 
 loops in the $gg$ induced signal processes
have been kept as independent parameters and their values varied between 
300 GeV and 1 TeV, the minimum figure being chosen in such a way that the
superpartners do not enter the $H^\pm$ decay chain, as for the upper mass of 
the latter we have taken the value of 600 GeV. Further notice that 
squark masses have been considered degenerate, for illustrative purposes,
so that only sbottom and stop loops in practice contribute.

In the numerical calculations presented in the next Section we have
adopted the following values for the electromagnetic coupling constant and
the weak mixing angle: $\alpha_{em}= 1/128$ and $\sin^2\theta_W=0.2320$.
The strong coupling constant $\alpha_s$, which appears in higher orders in
the computation of the charged Higgs decay widths and enters in some of
the production mechanisms, has been evaluated at one or two loops,
depending on the Parton Distributions Functions (PDFs) used, with
$\Lambda^{(4)}_{\overline{\mathrm {MS}}}$ (for the number of active
flavours $N_f=4$) input according to the values adopted in the fits of the
latter. For the leading-order (LO) package CTEQ4L \cite{CTEQ4}, which
constitutes our default set (as in \cite{BBK}), we have taken 236 MeV. The
factorisation/renormalisation scale $Q$ entering in both $\alpha_s$ and
the PDFs was set to $\sqrt{\hat{s}}$, the centre-of-mass (CM) energy at
the partonic level, in all processes generated.

In order to have an estimate of the dependence of the bottom quark and
gluon structure
function we tested our  signal rates against the next-to-leading
(NLO) sets MRS(R1,R2,R3,R4) \cite{mrs96}, i.e., the Martin-Roberts-Stirling
packages of the same `generation' as the CTEQ ones considered here,
plus the newly presented sets MRST \cite{mrst}, which embody new
data and an improved description of the gluons at small $x$, along
with a dedicated treatment of the heavy quark structure functions.
Typical differences were found to be within 15--20\%, 
at cross section level, with the shape of the relevant differential
distributions being little affected by the treatment of the partons
inside the proton. 

For the gauge boson masses and widths we have taken
$M_{Z}=91.1888$ GeV, $\Gamma_{Z}=2.5$ GeV,
$M_{W^\pm}=80.23$ GeV and
$\Gamma_{W^\pm}=2.08$ GeV. For the top mass we have used $m_t=175$ GeV
with the corresponding width evaluated at tree-level in the MSSM (yielding 
$\Gamma_t=1.55$ GeV if $M_{H^\pm}>m_t-m_b$, the Standard Model value). 
Bottom quarks have been considered massless
when treated as partons inside the proton, while a finite value of 4.25
GeV (pole mass) has been retained in the final states. Note that in
calculating the ME for the decay process $H^\pm\ar tb$ the  
Yukawa mass of the $b$ quark has been run up to the charged Higgs mass scale, 
in accordance to the way the corresponding width has been computed.
Finally, for simplicity, we set the Cabibbo-Kobayashi-Maskawa
matrix element $V^{bt}_{\mathrm{CKM}}$ to one.

\section*{3. Results}

As it is impractical to cover all possible regions of the MSSM parameter
space $(\MA,\tb)$, we concentrate here on the two representative (and
extreme) values $\tan\beta=1.5$ and 30, and on masses of the charged
Higgs boson in the range 160 GeV $\Ord\MHpm\Ord$ 600 GeV. The large
bibliography existing on the MSSM Higgs decay phenomenology should allow
one to easily extrapolate our results to other values of \tb\
\cite{guide}.

We consider the semi-leptonic event modes for processes
(\ref{bb})--(\ref{tt}):
\begin{equation}\label{signature}
W^\pm H^\mp\ar W^\pm tb\ar b\bar b W^+W^-\ar b\bar b~~jj~~\ell 
+ \mbox{missing~energy/momentum}.
\ee

In fact, we base our signal selection procedure on the following 
general requirements.
\begin{enumerate}
\item High purity double $b$ quark tagging. This may be expected to yield
an efficiency of at least 50\% per fiducial $b$ jet \cite{ATLAS,CMS}. This is
essential considering the large rate of $W^\pm+~\mbox{jet}$ events with
light quarks and gluons. All our results will
assume 100\% bottom quark tagging efficiency $\epsilon_b$, and thus will 
eventually need to be multiplied by the actual $\epsilon_b^2$ once we will 
have the LHC detectors running.
\item Lepton isolation at high transverse momentum. Selecting semi-leptonic
events should enable one to use the high $p_T$ and
isolated lepton (electron and/or muon) originating from the $W^\mp$ produced 
in association with the $H^\mp$ as a clean trigger\footnote{For the 
time being, we neglect $W^\pm\ar \tau\nu_\tau$ decays, which should also 
be identified easily thanks to their `one-prong' signatures, as remarked in
\cite{BBK}.} \cite{BBK}. In addition, the light quark jets $j$ coming
from the secondary $W^\mp$, from $H^\mp\ar tb\ar b\bar b W^\mp$, would
allow for the reconstruction of the Higgs mass peak. However, one should 
recall that the two gauge bosons could decay the other way round, this
in principle spoiling the efficiency of the signal selection.
In practice, one can remove the contribution from semi-leptonic Higgs
decays by simply imposing cuts on the reconstructed top mass.
\item $W^\pm$ and $t$ mass reconstruction in two and three jet combinations,
respectively, to eliminate QCD multi-jet production.
\end{enumerate}

Our results are shown throughout Figs.~\ref{fig:xsections}--\ref{fig:hist}. 
When differential spectra are plotted, the three 
representative parameter space points of $\MA=200$, 400 and 600 GeV 
at $\tbb$ have been chosen, corresponding to $\MHpm=214$, 407 and 605 GeV,
respectively.

As a preliminary exercise, in order to understand the kinematics of the
signal and background better, and possibly to pin down systematic
differences which can be used in the selection of candidate Higgs events,
we compare their total and differential rates in the channel (\ref{signature})
without the usual detector
cuts on transverse momenta and pseudorapidity. In fact, both processes
have finite production rates at lowest order over all of the phase space.

The main frame of Fig.~\ref{fig:xsections} shows the signal (with no squark 
loop contributions) and background cross
sections before any detector or selection cuts, evaluated at the LHC energy
(14 TeV) and plotted against the $H^\pm$ mass  for the two values of
$\tan\beta$. The background remains constant as a function of the charged
Higgs mass, except when the decay $t\ar bH^+$ is kinematically allowed. It
is found that even when the signal cross section is the highest the
signal/background ratio is less than one in a thousand. Notice that
at decay level,  see eq.~(\ref{signature}), the signal rates suffer from a 
further small depletion (in addition to that due to finite width effects)
as compared to the on-shell ones of Ref.~\cite{BBK}, if
the latter are multiplied by the relevant branching ratios (BRs). 
This is due to the fact that the Goldstone part
of the longitudinal component of the $W^\pm$ boson 
(i.e., the wave-function term proportional to 
$p_{W^\pm}^\mu/M_{W^\pm}$) produced in association 
with the $H^\mp$ scalar in $b\bar b$ fusion
(included in that reference) does not survive the decay into `massless' 
fermions. The effect depends on $\tb$, owning 
the structure of the MSSM couplings in the $\Phi W^\pm H^\mp$ vertices, where
$\Phi=H$ and $h$, and on the relative strength of the two 
contributing diagrams (see Figs.~1--2 of Ref.~\cite{BBK}).

In the central insert of Fig.~\ref{fig:xsections} we study possible virtual
effects of SUSY, manifesting itself in the triangle diagrams of $gg$ fusion.
In fact, we plot the ratio of the signal cross sections obtained by adding 
the rates of
both suprocesses (\ref{bb}) and (\ref{gg}), the latter including squark loops,
against those calculated when such contributions are neglected.
As already remarked in the literature \cite{SDGZ},
sizable SUSY effects in the $gg$ subprocess are expected only for
squark masses below 500 GeV or so and particularly at small $\tb$.
Thus, in our plot we present the ratios for $m_{\tilde q}=300$ and 500 
GeV at $\tba$ only. Given the remarks made in the Introduction, concerning
our remotion of the box diagrams and, consequently, of the cancellations
against the triangle ones (which we would further think to be active for 
virtual squark contributions as well), our numbers should in this 
circumstance be intended as a sort of {\sl upper limit} that one might 
expect from such SUSY effects. From this prospect, it is then clear 
that production rates can increase by no more than 20\% or so, and provided
squark masses are rather low, a correction indeed 
comparable to the combined uncertainties related to the $b$ quark and gluon 
PDFs  in (\ref{bb}) and (\ref{gg}), respectively, and to the scale dependence 
of the $gg$ production rates \cite{SDGZ}. For this reasons, and to simplify
the discussion as well, hereafter, we will neglect altogether 
the squark loop contributions in our signal rates.

The differential spectra are displayed in Figs.~\ref{fig:pt}--\ref{fig:mblj}.
In particular, we plot\footnote{We make no distinction 
between $b$ and $\bar b$ jets, tacitly
assuming neither jet charge determination nor lepton tag.} 
\begin{itemize}
\item the lepton, light quark, $b$ jet and missing transverse momentum;
\item the lepton, light quark and $b$ jet pseudorapidity;
\item the lepton/light quark jet and lepton/$b$ jet separation,
defined by the variable $\Delta R=\sqrt{(\Delta\eta)^2+(\Delta\phi)^2}$
in terms of pseudorapidity $\eta$ and azimuth $\phi$.
\end{itemize}
Furthermore, we present the invariant mass spectra of the following systems:
\begin{itemize}
\item $bb$, as obtained by pairing the two jets with displaced 
vertices;
\item three jets, with only one $b$ jet involved;
\item four jets, involving all four jets in the final state.
\end{itemize}
The combinatorics in the three, four and lepton/light quark jet systems is 
accounted for by simply plotting all possible momentum combinations each with
the same event weight. In other terms, the signal spectra contain both 
hadronic and leptonic $W^\pm/H^\mp$ decay modes (but not their interference, 
that we expect negligible), which have in fact different kinematics. 
The two component will eventually be separated by imposing cuts around the 
reconstructed top quark mass, as mentioned before. In contrast, the background 
spectra, obviously, do not depend on whether the top or the antitop decays 
leptonically.

The usual distributions in $p_T$ and $\eta$ indicate the effects of
detector acceptance cuts on the signal and background samples.
Neither of these affects the event rates significantly. The distribution
in $\Delta R$ indicates that the requirement of lepton/jet separation will not
harm the event rates either. Furthermore, 
the signal and background distributions are
very similar and it is clear that none of these variables 
can profitably be used to optimise the
 selection procedure. Incidentally, we mention that we 
also had a look at the transverse momentum and pseudorapidity of the three 
hadronic systems introduced above, without finding any significant 
difference between Higgs and top events.

Presumably, the invariant mass distributions,
see Fig.~\ref{fig:m} and \ref{fig:mblj} for the purely hadronic and
semi-leptonic systems, respectively, will give us the greatest
chance of removing the background. By imposing cuts on the two light quark 
jet and two light quark plus bottom jet invariant masses around the $W^\pm$
and top quark resonances, respectively, we can remove most of the QCD noise,
having to deal finally with the semi-leptonic top pair decays which is the
greatest challenge. As a matter of fact, di-jet pairs of light 
quarks from $t\bar t$ events naturally peak at $M_{W^\pm}$, exactly
as those from  $W^\pm H^\mp$ do. As for the three jet systems,
mispairings of $b$ quarks with the wrong $W^\pm$ have more severe effects
on the signal than on the background, as one can intuitively expect
from the production dynamics and as it can be appreciated in the middle
plot of Fig.~\ref{fig:m} (note the height of the peak for $t\bar t$ events,
as compared to that of the $W^\pm H^\mp$ ones).

Here we propose the following related and complementary cuts
on invariant masses. For a start,
we put ourselves in the favourable phenomenological position that the
charged Higgs mass $M_{H^\pm}$ is known, e.g., thanks to a previous
detection of the light scalar Higgs $h$ and  to the measurement of its
couplings. Under these circumstances, in order to enhance the $W^\pm
H^\mp$ to $t\bar t$ rates, one can impose a cut on the invariant mass of
the $bb$ pair. Since in the signal both bottom jets originate from the $H^\pm$
scalar 
and assuming that $H^\pm\ar b\bar b W^\pm$, the invariant mass squared
$M_{bb}$ must be below $\sqrt{M^2_{H^\pm}-M^2_{W^\pm}}$ (apart from finite 
width effects). The distribution
from top pair events has a characteristic scale of $2m_t$ and therefore, at
low $M_{H^\pm}$, it can be filtered out by setting a sufficiently low cut.
The gain for the signal-to-background rate is large if $H^\pm$ is
reasonably ($m_t<M_{H^\pm}<2m_t$) light (solid and dashed curves in
Fig.~\ref{fig:m}, upper plot), whereas for heavy charged Higgs bosons, when
$M_{H^\pm}$ is of the order $2m_t$ or greater, the cut is not useful
(dotted curve in Fig.~\ref{fig:m}, upper plot). However, one should note that 
this selection cut can be utilised
successfully only when $M_{H^\pm}$ is approximately known, and is of
limited use even then, since it only removes  the background from
regions of phase space far away from reconstructing the charged Higgs mass
peak.

Similarly, one can impose a cut on the invariant mass of a $b\ell$ pair,
where $b$ is the bottom jet which does not reproduce the top quark with
the light di-jet pair (i.e., the one yielding the reconstructed
$m_t$ further away from its actual value). Here, the selection works
because for the $t\bar t$ background, if both top and $W^\pm$ are on the
mass shell, one has $M_{b\ell}<\sqrt{m^2_t-M^2_{W^\pm}}$. However, it
should be noticed that the Higgs production mechanism can really push the
$M_{b\ell}$ value beyond $\sqrt{m^2_t-M^2_{W^\pm}}$ only if $\MHpm$ is
large enough: see dashed and dotted curves in Fig.~\ref{fig:mblj}, 
for $M_{b\ell/j}\OOrd160$ GeV. Failing this condition, the suppression 
against the signal itself can be quite large (e.g., an additional rejection 
factor of five for $\MHpm=214$ GeV at $\tbb$). However, it turns out that
such a constraint is definitely necessary to bring down the  background
rates to manageable levels (as it contributes with an additional factor of
thirty or so to the suppression of top-antitop events), so that we employ
it even at low $\MHpm$ values.

In addition, we have made the following, more standard cuts:
\begin{enumerate}
\item isolation of the two bottom jets from the light quark jets and
from each other, as we tentatively set the azimuthal-pseudorapidity
separation at $\Delta R_{bb,bj}>0.7$ between them\footnote{Note that we
allow for the light quark jets to be arbitrarily close.};
\item for an isolated lepton, we
impose the cut $\Delta R_{\ell b,\ell j}>0.4$ between the lepton and all jets;
\item the light quark jet pair mass $M_{jj}$ within $M_{W^\pm}\pm10$ GeV;
\item the light quark jet pair and a bottom quark jet combine at least once 
to a mass $M_{bjj}$ of $m_t\pm10$ GeV;
\item all one-particle pseudorapidities (of leptons, light and heavy quarks) 
are constrained within a detector region of 2.5;
\item the transverse momentum cut was set at 20 GeV for all jets and
leptons, and for the missing transverse momentum as well.
\end{enumerate}

Fig.~\ref{fig:xsections_cuts} displays the total 
rates for signal and background
after the above selection cuts have been enforced. Since the latter depends
on the value of $\MHpm$, the cross section for events of the type (\ref{tt})
is no longer constant when $M_{H^\pm}\OOrd m_t$. However, given the weak 
dependence 
of $\MHpm$ on $\tb$, the two top-antitop curves overlap in that mass range.
Although the signal-to-background ratio has greatly improved, as compared
to the initial situation in Fig.~\ref{fig:xsections},  for any 
combination of $\MHpm$ and $\tan\beta$, this is still very small, at least
one part in a hundred, so to presumably dash away any hopes of resolving
the Higgs peak.

In fact, to be realistic, one should expect a four jet mass resolution
of no less than 10 GeV, given the usual uncertainties in reconstructing
parton directions and energies from multi-hadronic events. Therefore, in
Fig.~\ref{fig:hist}, we have binned the invariant masses of the 
$bbjj$ system in signal and background using that value. If one does so,
it is clear that the Higgs mass peaks at 214, 407 and 605 GeV (with 
Breit-Wigner width $\Gamma_{H^\pm}$ of 1.2, 4.4 and 6.1 GeV, respectively),
at $\tbb$, are overwhelmed by the top-antitop events. Seemingly, even where
the signal is more pronounced over the background (at large $\MHpm$),
the excess amounts to no more than 10\% at the most in the central bin. This
is probably too little, further considering, on the one hand, the 
aforementioned
uncertainties (PDFs, $K$-factors, etc.) and, on the other hand, that the 
event rate is poor, only around 1 fb prior to the vertex tagging 
efficiency $\epsilon_b^2$ being applied. Needless to say, if one looks back at 
Fig.~\ref{fig:xsections_cuts},
similar conclusions should be expected for all other combinations of
$M_A$ and $\tan\beta$ considered here, finally recalling our systematic
overestimate of the signal rates for low values of the latter.

\section*{4. Conclusions}

We believe that in the $H^\mp\ar tb\ar b\bar b W^\mp$ channel,
{\sl heavy} charged Higgs scalars of the 
Minimal Supersymmetric Standard Model produced in association with 
$W^\pm$ gauge bosons cannot be resolved at the LHC,
via semi-leptonic $W^+W^-$ decays,  
for Higgs masses in the range between $2m_t$ and 600 GeV (those producible at 
observable rate), at neither low nor high $\tb$, because of the presence of 
the irreducible background from top-antitop events. Furthermore,
our results can safely be applied to a more general 2 Higgs Double Model too
(where  mass and coupling constraints in the Higgs sector can be relaxed), 
given the extremely poor significance of the $W^\pm H^\mp$ rates over the
$t\bar t$ ones. As for other hadronic collider environments, the prospects
of detection at the Tevatron (Run II) are even more reduced, given the
lower machine luminosity and since the production cross
section of the signal is there about three orders of magnitude smaller
than at the CERN accelerator (for detectable Higgs masses below 300 GeV or so),
while the background only decreases by about two orders.

We have reached these conclusions 
after performing a detailed signal-to-background
analysis, based on matrix element calculations of elementary 2 $\ar$ 6
suprocesses, convoluted with up-to-date parton 
distribution functions, and exploiting dedicated selection cuts, beyond
the usual requirements in transverse momentum and pseudorapidity. Although we 
have confined ourselves to the the parton level only, wherein jets are 
identified with partons, we are however 
confident that hadronisation and detector effects will not modify our main 
results.

Nonetheless, we would like to conclude this paper with a positive note.
Charged Higgs production in association with $W^\pm$'s, 
via $b\bar b$ and $gg$ fusion at hadron colliders, represents a novel
mechanism, whose decay phenomenology is largely unknown and that
should be investigated further, considering that the detection of this 
particle is {\sl not at all} certain
at the next generation of hadronic machines, especially in the heavy mass
range. In this respect, we would like to advocate, for example, 
the consideration of non-Standard Model decay channels, involving squarks, 
sleptons and gauginos, which was beyond the intention of this study.

\subsection*{Acknowledgements}

SM is grateful to the UK PPARC and
KO to Trinity College and the Committee of Vice-Chancellors and 
Principals of the Universities of the United Kingdom for
financial support. SM also thanks David Miller and Mike Seymour for
useful discussions.

\goodbreak

\vfill

\clearpage\thispagestyle{empty}
\begin{figure}[p]
~\epsfig{file=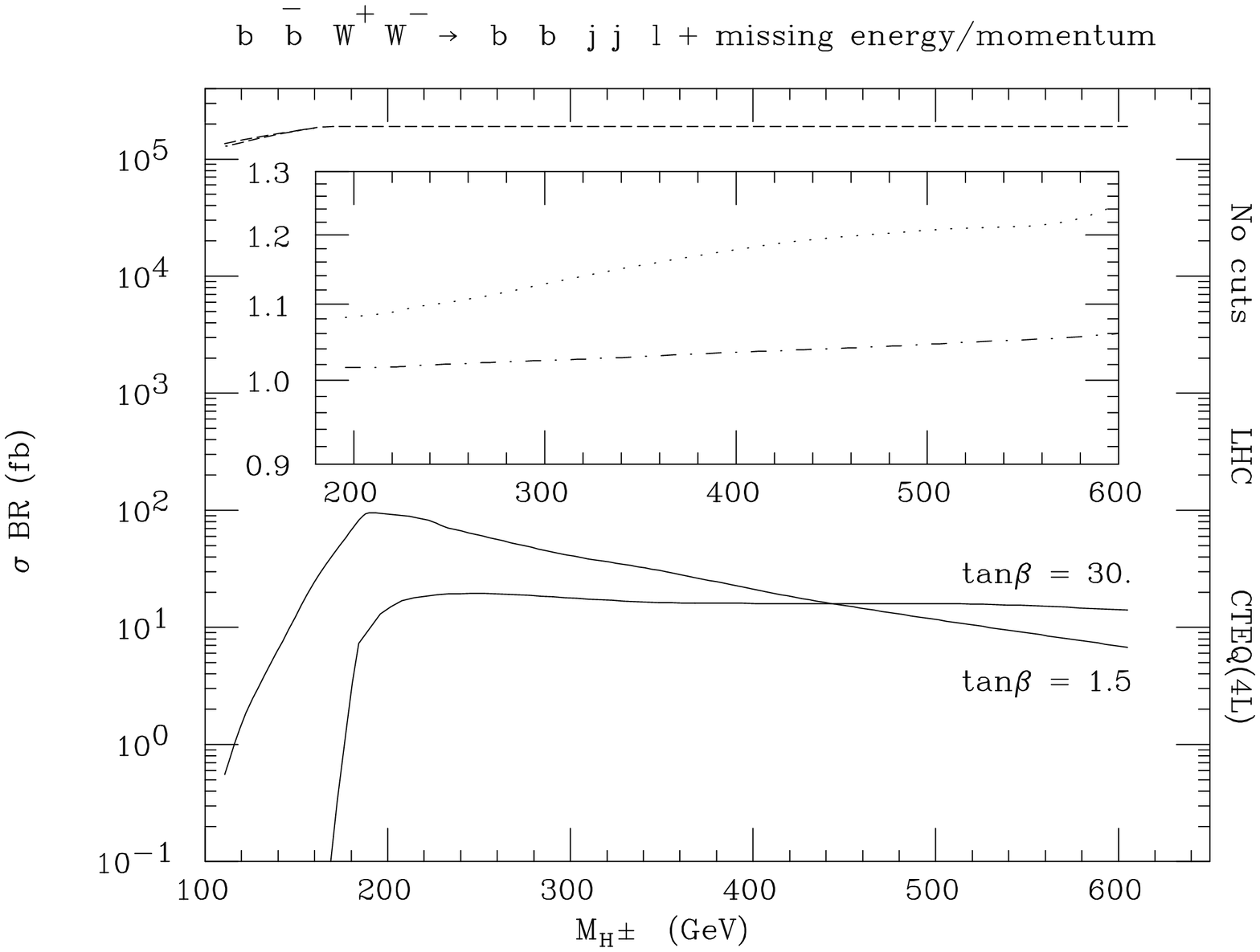,height=16cm,angle=90}
\vspace*{-1cm}
\caption{Total event rates  for $W^\pm H^\mp$ (solid) and $t\bar t$ production
(dashed) at the LHC, with no cuts implemented, using CTEQ(4L), 
as a function of $\MHpm$ for $\tba$ and $\tbb$.
In the blow up figure, the $W^\pm H^\mp$ production rates  for $\tba$ including
squark loop contributions, with $m_{\tilde q}=300$ (dotted) and 500 
(dot-dashed) GeV, divided by those obtained when the latter are neglected. 
For $\tbb$, squark contributions are negligible, so that the same ratios  
would visually coincide with one.}
\label{fig:xsections}
\end{figure}

\clearpage\thispagestyle{empty}
\begin{figure}[p]
~\epsfig{file=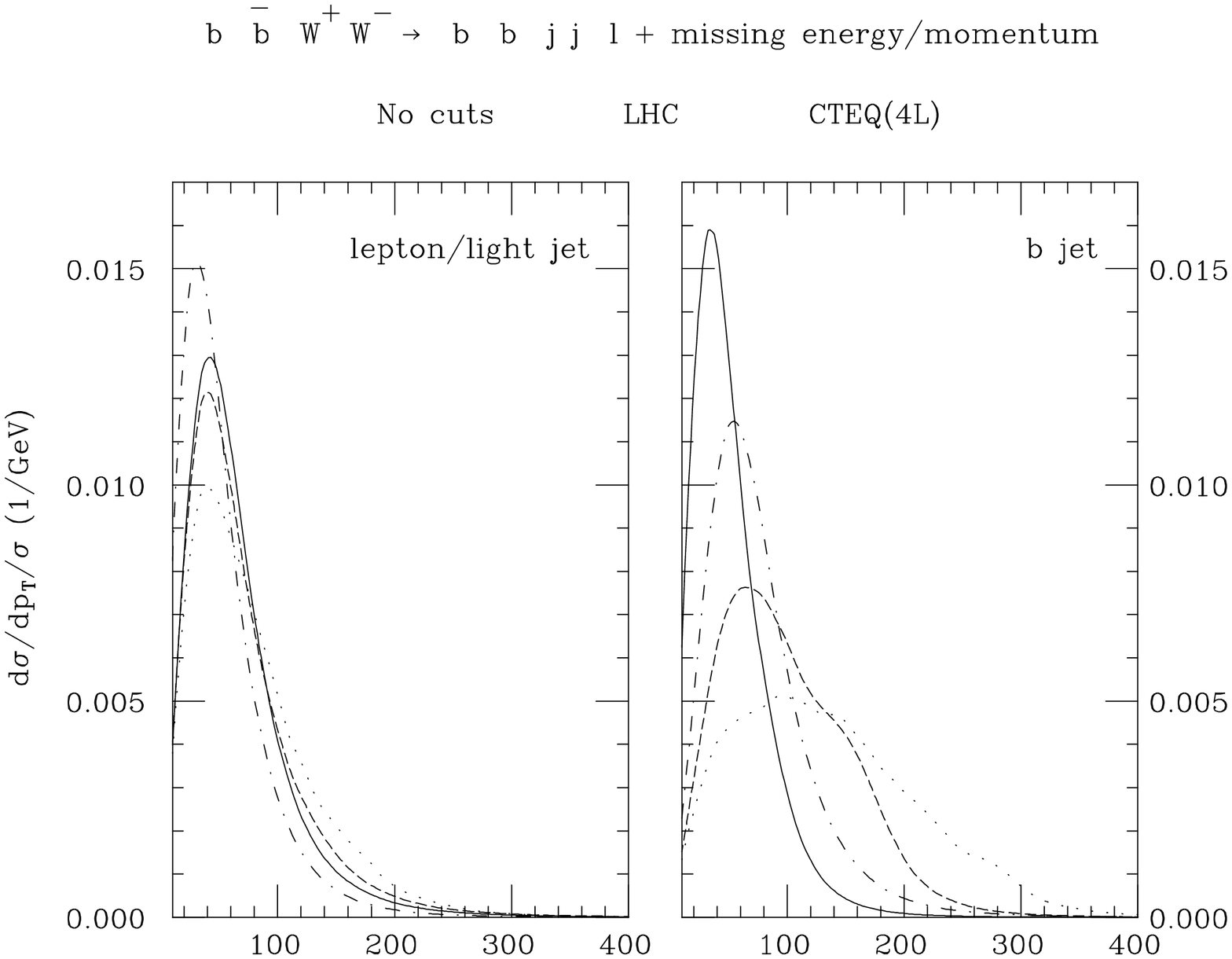,height=16cm,angle=90}
\vskip-3.5cm
~\epsfig{file=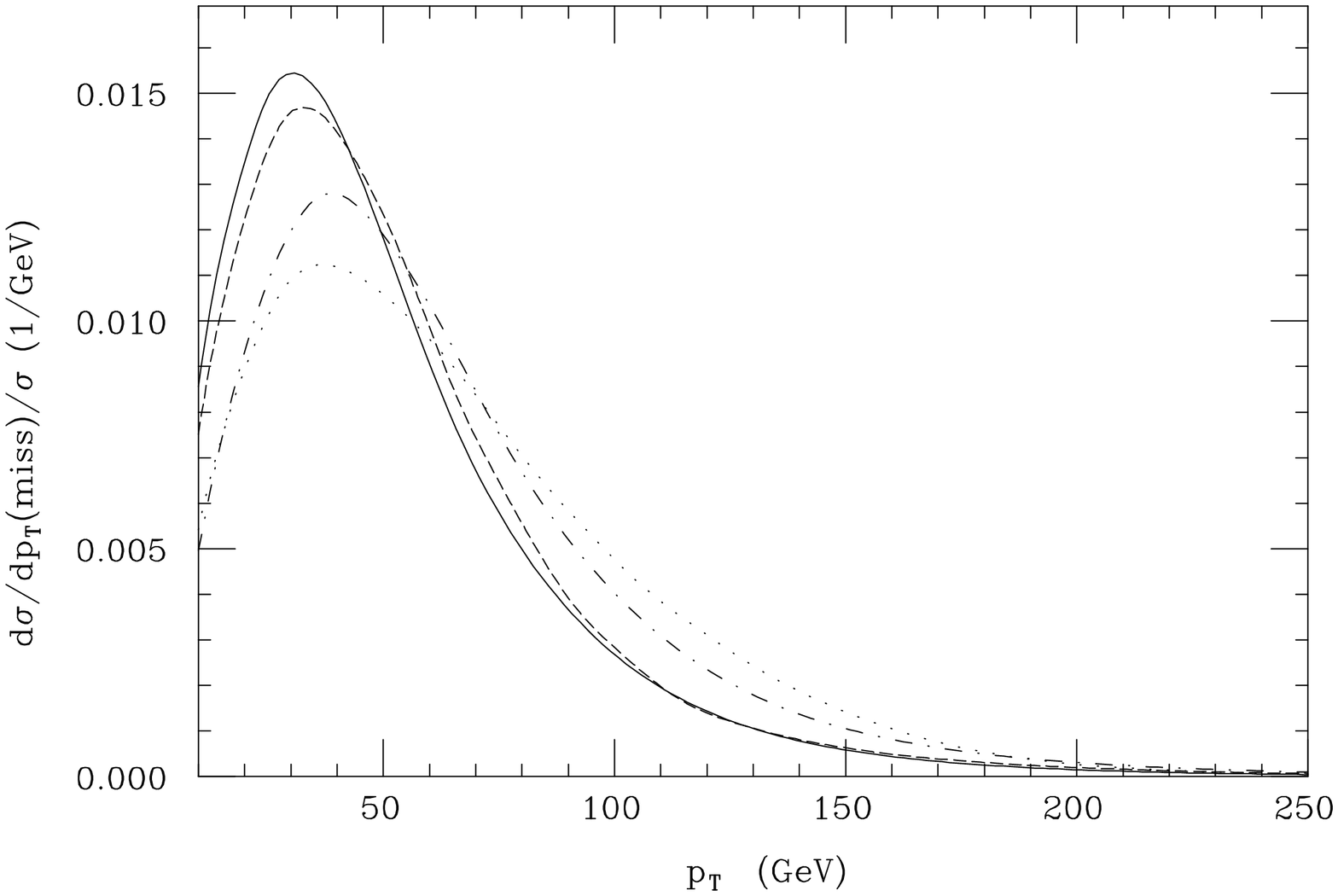,height=16cm,angle=90}
\vspace*{-1.0cm}
\caption{Differential spectra in
lepton/light quark jet (top-left),  $b$ jet (top-right) and missing (bottom)
transverse momentum.
Here, $\tbb$ and the $W^\pm H^\mp$ rates are
plotted for $\MHpm=214$ (solid), 407 (dashed) and 605 (dotted) GeV. 
As $\MHpm>m_t+m_b$, $t\bar t$ events have no MSSM parameter dependence 
(dot-dashed). No cuts have been implemented. The PDFs CTEQ(4L) have been used. 
Distributions are normalised to unity. Note that, in the case of 
the lepton and light quark jet, the curves coincide.}
\label{fig:pt}
\end{figure}

\clearpage\thispagestyle{empty}
\begin{figure}[p]
~\epsfig{file=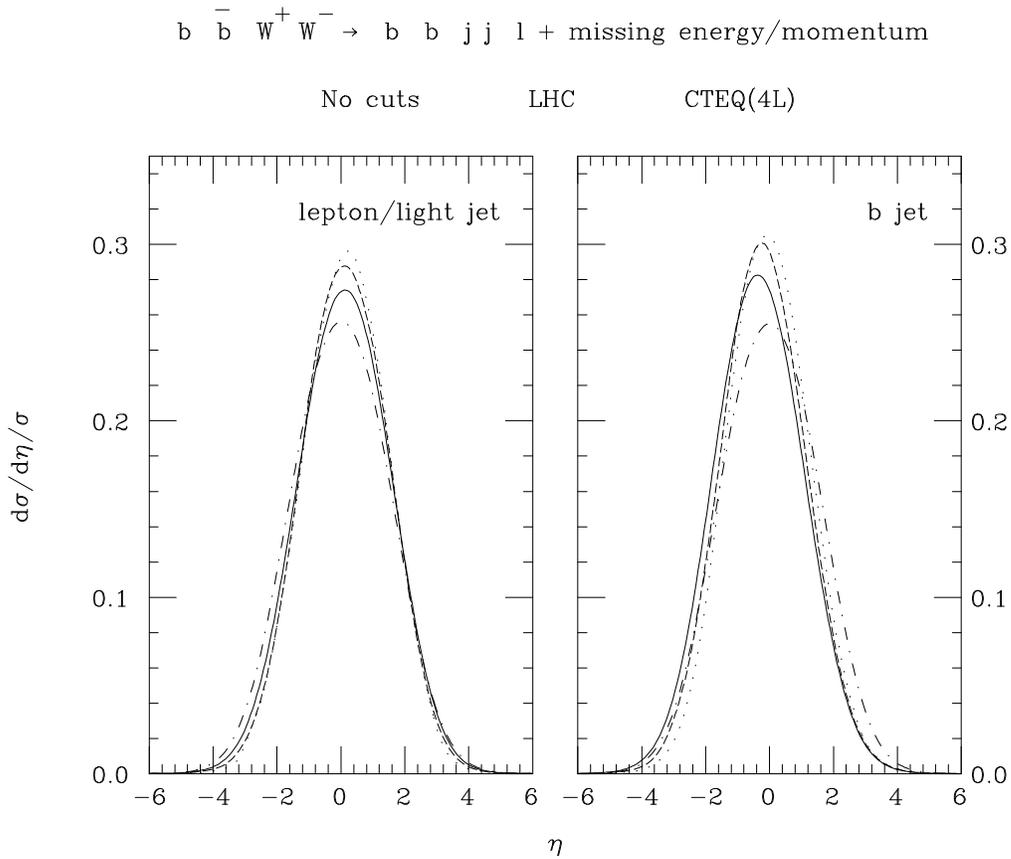,height=16cm,angle=90}
\caption{Differential spectra in
lepton/light quark jet (left) and  $b$ jet (right)
pseudorapidity. Here, $\tbb$ and the $W^\pm H^\mp$ rates are
plotted for $\MHpm=214$ (solid), 407 (dashed) and 605 (dotted) GeV. 
As $\MHpm>m_t+m_b$, $t\bar t$ events have no MSSM parameter dependence 
(dot-dashed). No cuts have been implemented. The PDFs CTEQ(4L) have been used. 
Distributions are normalised to unity. Note that, in the case of 
the lepton and light quark jet, the curves coincide.}
\label{fig:eta}
\end{figure}

\clearpage\thispagestyle{empty}
\begin{figure}[p]
~\epsfig{file=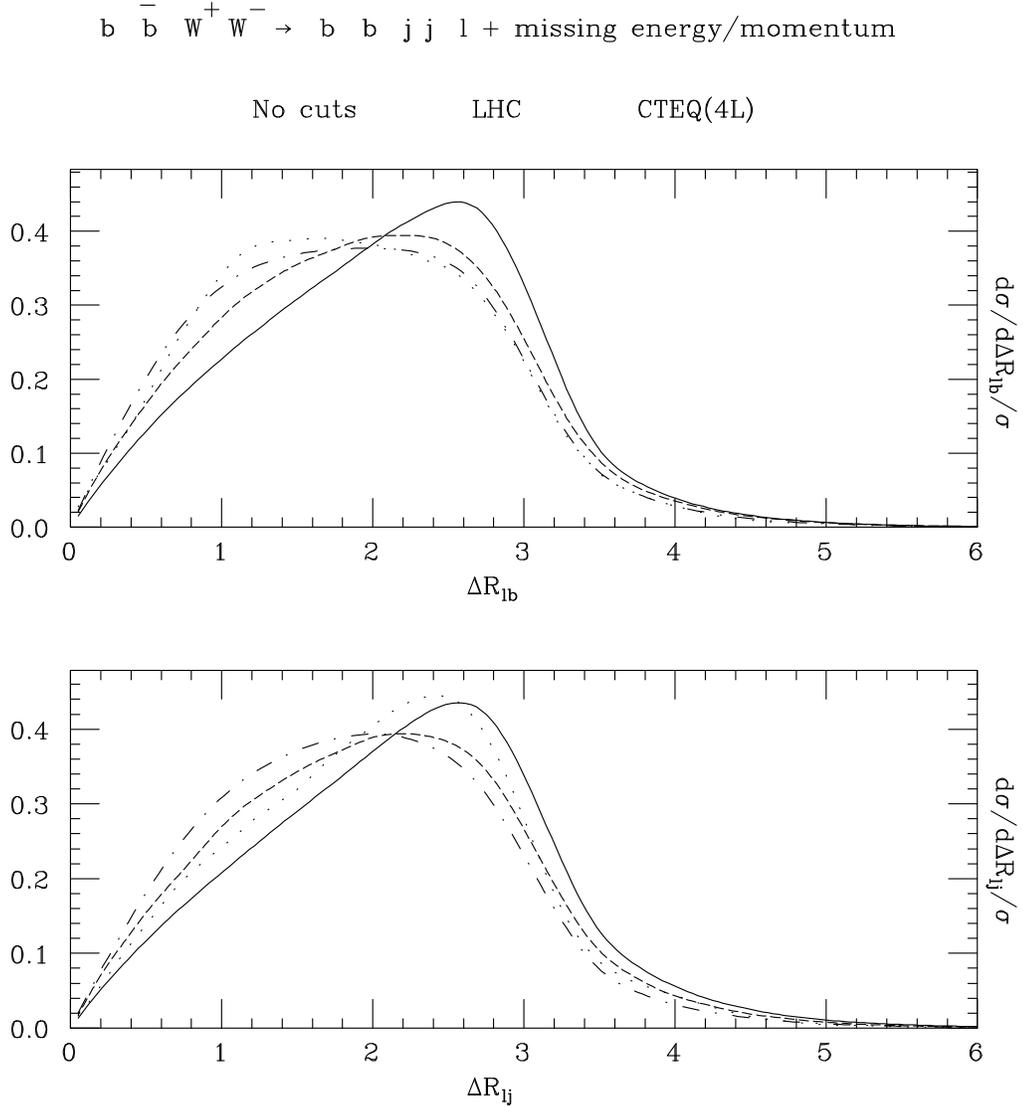,height=22cm,angle=0}
\vspace*{-4cm}
\caption{Differential spectra in pseudorapidity-azimuth 
separation between the 
following pairs of particles in $W^\pm H^\mp$ and $t\bar 
t$ events: lepton/$b$ jet (top); lepton/light quark jet (bottom). 
Here, $\tbb$ and the $W^\pm H^\mp$ rates are
plotted for $\MHpm=214$ (solid), 407 (dashed) and 605 (dotted) GeV. 
As $\MHpm>m_t+m_b$, $t\bar t$ events have no MSSM parameter dependence 
(dot-dashed). No cuts have been implemented. The PDFs CTEQ(4L) have been used. 
Distributions are normalised to unity.}
\label{fig:r}
\end{figure}

\clearpage\thispagestyle{empty}
\begin{figure}[p]
~\epsfig{file=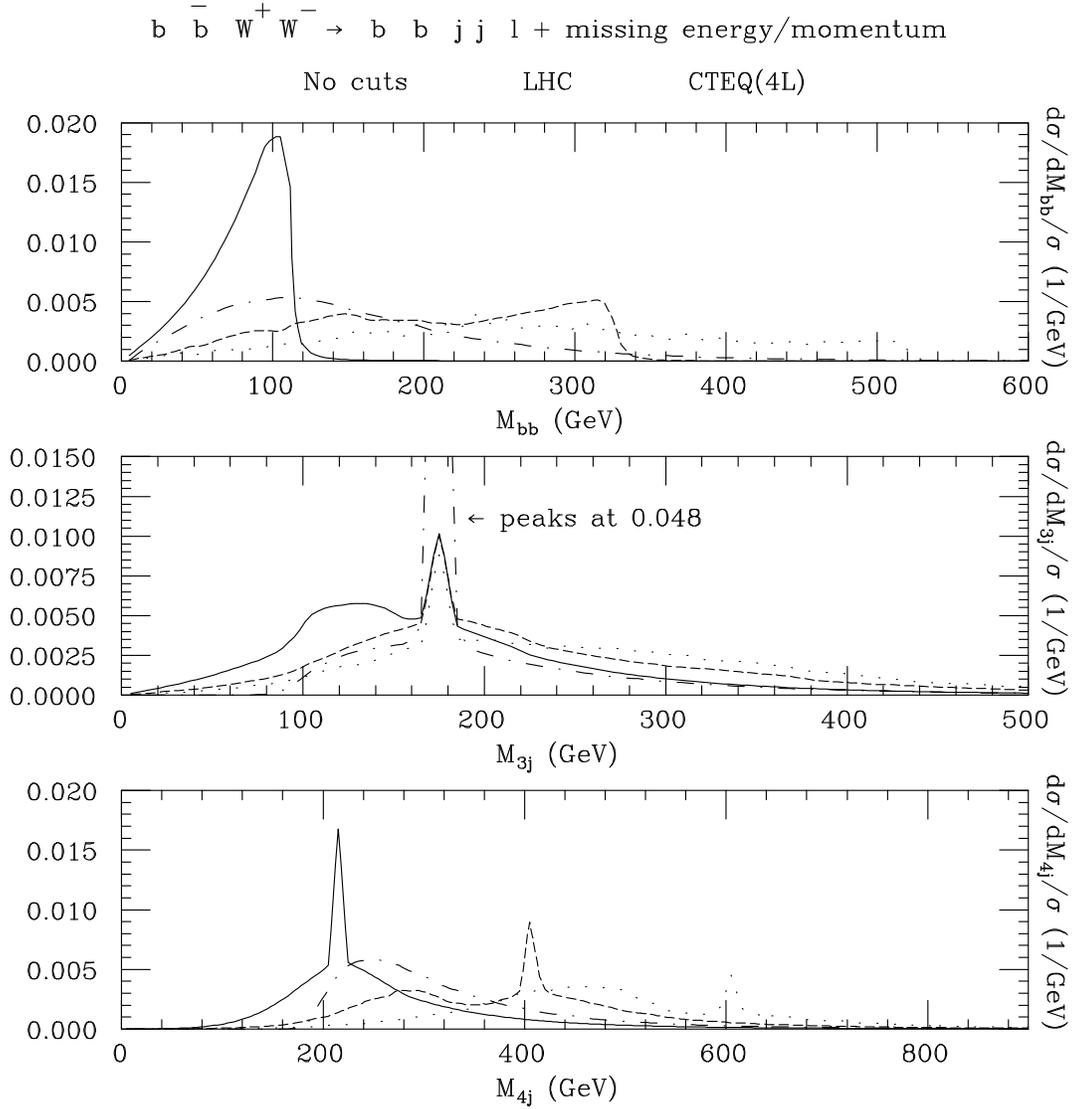,height=22cm,angle=0}
\vspace*{-4cm}
\caption{Differential spectra in invariant mass for the following
systems in $W^\pm H^\mp$ and $t\bar 
t$ events: $bb$ (top); three jet (middle); four jet (bottom). 
Here, $\tbb$ and the $W^\pm H^\mp$ rates are
plotted for $\MHpm=214$ (solid), 407 (dashed) and 605 (dotted) GeV. 
As $\MHpm>m_t+m_b$, $t\bar t$ events have no MSSM parameter dependence 
(dot-dashed). No cuts have been implemented. The PDFs CTEQ(4L) have been used. 
Distributions are normalised to unity.}
\label{fig:m}
\end{figure}

\clearpage\thispagestyle{empty}
\begin{figure}[p]
~\epsfig{file=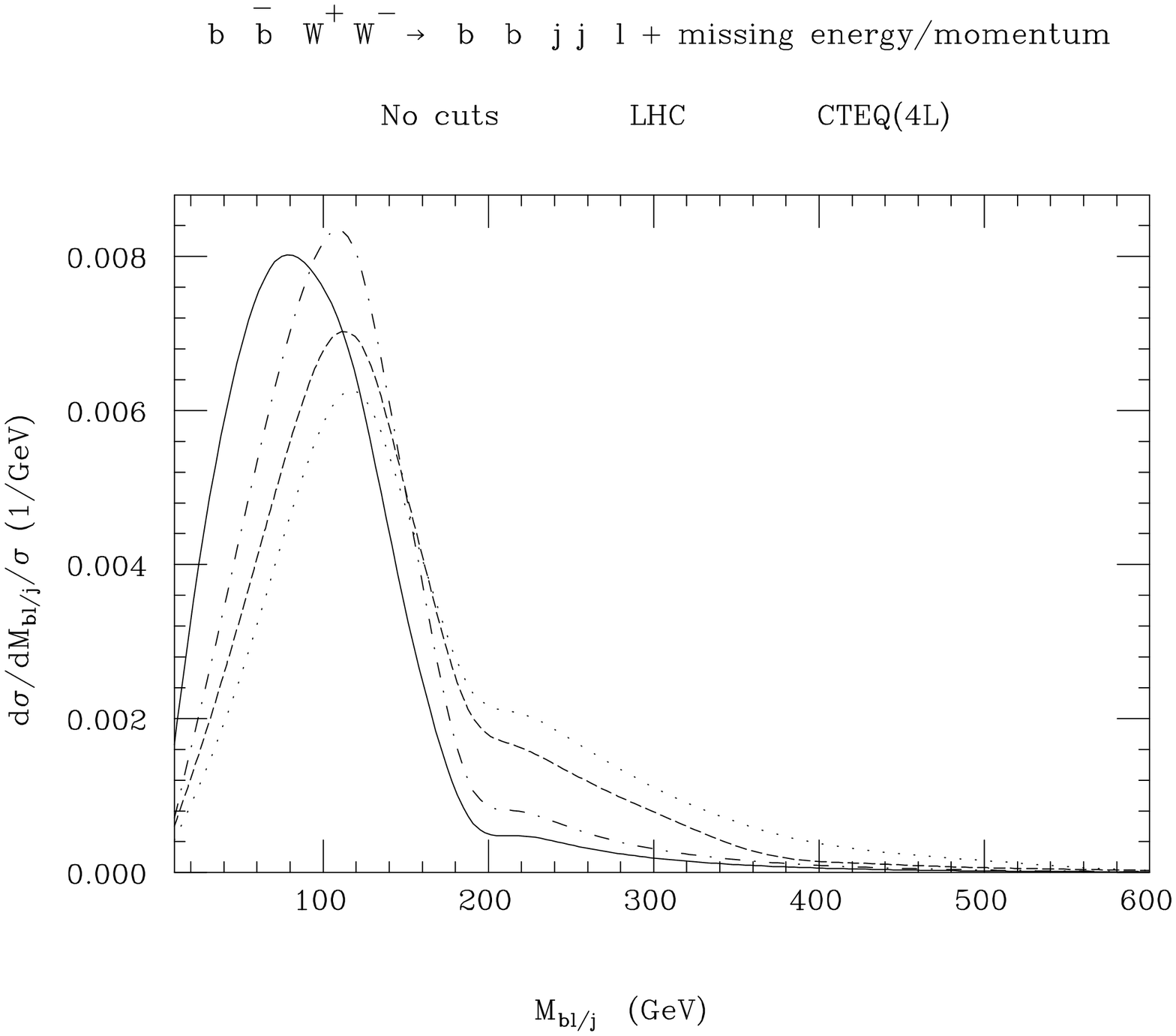,height=16cm,angle=90}
\caption{Differential spectra in invariant mass for the $b\ell/\mbox{jet}$
system in $W^\pm H^\mp$ and $t\bar t$ events.
Here, $\tbb$ and the $W^\pm H^\mp$ rates are
plotted for $\MHpm=214$ (solid), 407 (dashed) and 605 (dotted) GeV. 
As $\MHpm>m_t+m_b$, $t\bar t$ events have no MSSM parameter dependence
(dot-dashed). No cuts have been implemented. The PDFs CTEQ(4L) have been used. 
Distributions are normalised to unity.}
\label{fig:mblj}
\end{figure}

\clearpage\thispagestyle{empty}
\begin{figure}[p]
~\epsfig{file=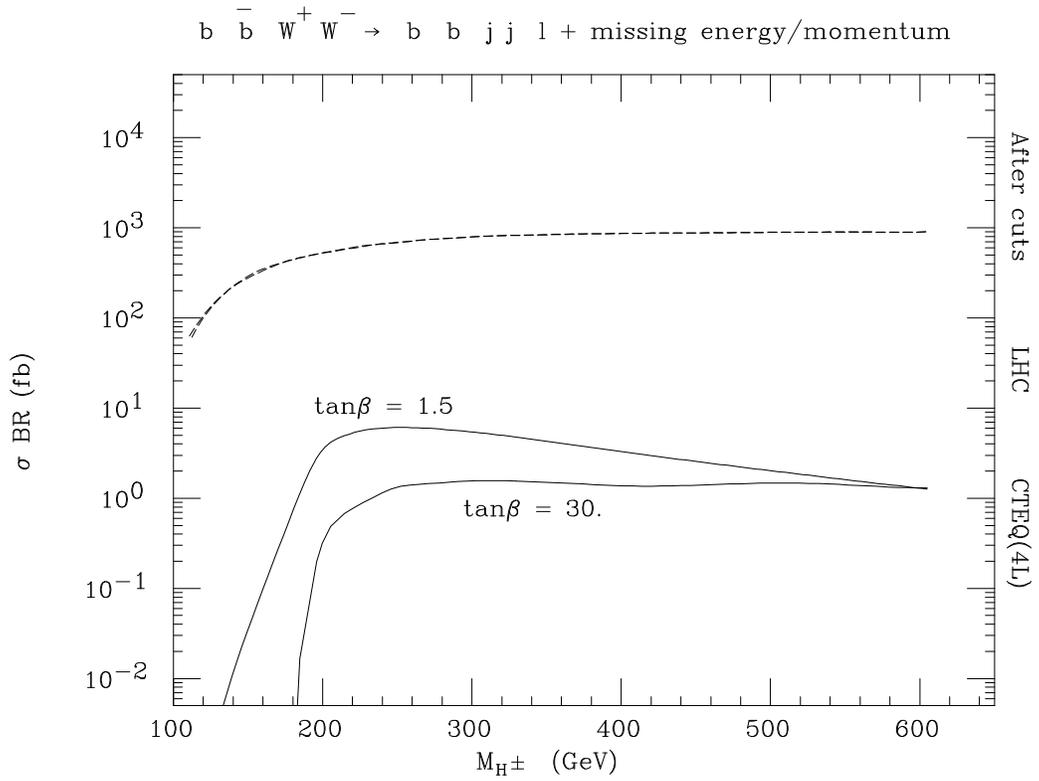,height=16cm,angle=90}
\vspace*{-1cm}
\caption{Total event rates  for $W^\pm H^\mp$ (solid) and $t\bar t$ production
(dashed) 
at the LHC, after the selection cuts have been implemented, using CTEQ(4L), 
as a function of $\MHpm$ for $\tba$ and $\tbb$.}
\label{fig:xsections_cuts}
\end{figure}

\clearpage\thispagestyle{empty}
\begin{figure}[p]
~\epsfig{file=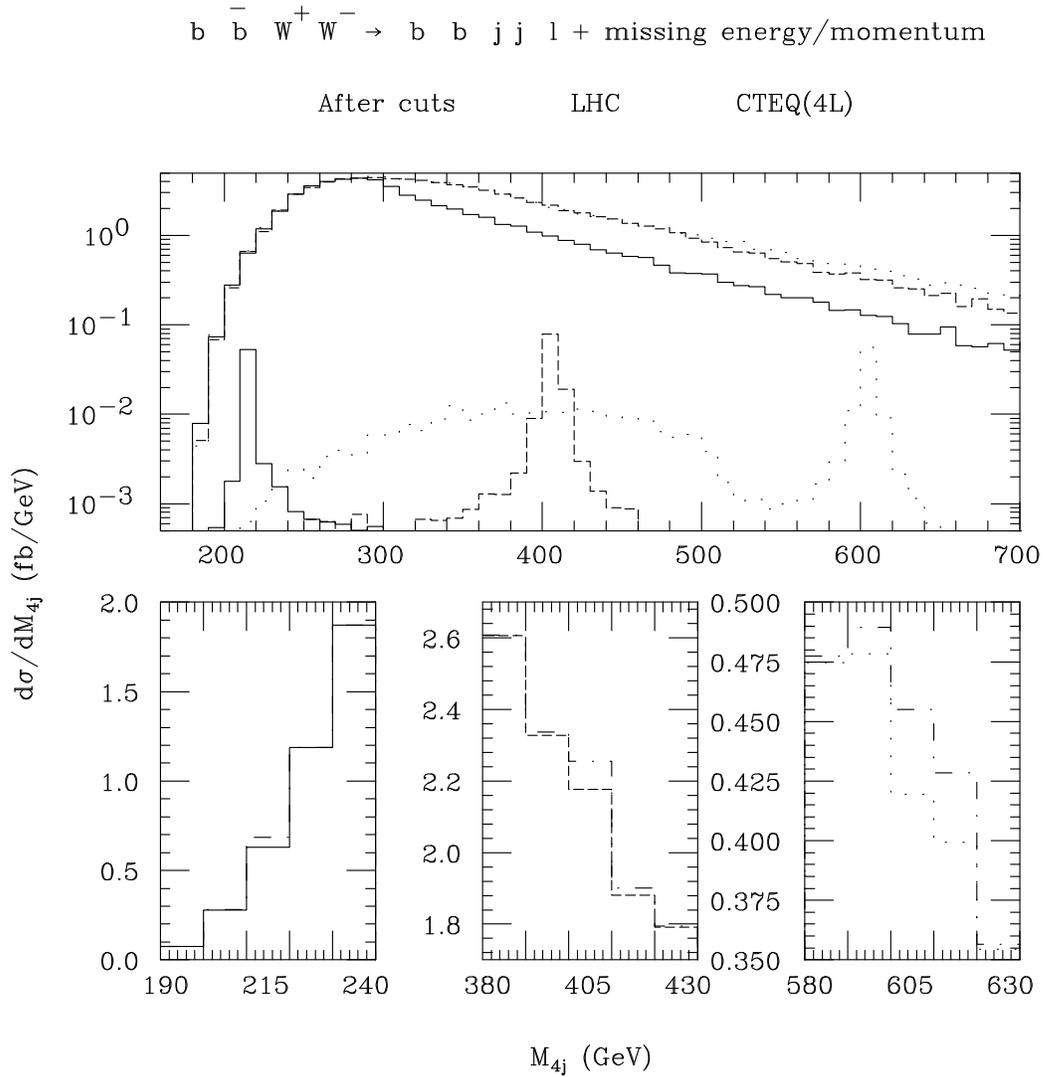,height=22cm,angle=0}
\vspace*{-3cm}
\caption{Differential spectra in the four 
jet invariant mass in $W^\pm H^\mp$ and $t\bar 
t$ events for $\tbb$ and $\MHpm=214$ (solid), 407 (dashed) and 
605 (dotted) GeV. Their sum (dot-dashed) is also reported in the three blow
up frames below, compared against  the $t\bar t$ rates from above.
Even though $\MHpm>m_t+m_b$, $t\bar t$ events have a MSSM parameter dependence
due to an $\MHpm$ based constraint being implemented in the 
selection cuts. The PDFs CTEQ(4L) have been used. 
Distributions are normalised to total cross sections.}
\label{fig:hist}
\end{figure}

\end{document}